\documentclass[aps,prb,twocolumn,superscriptaddress,showpacs]{revtex4}

\usepackage{amssymb}
\usepackage{graphicx}
\usepackage{natbib}
\usepackage[usenames]{color}
\usepackage[dvipsnames]{xcolor}

\begin{document}

\title{Electronic and ionic conductivities in superionic Li$_4$C$_{60}$}

\author{D. Quintavalle}
\affiliation{Budapest University of Technology and Economics, Institute of Physics and Condensed Matter Physics Research Group of the Hungarian Academy of Sciences, H-1521, Budapest P.O. Box 91, Hungary}
\thanks{Present address: Semilab Semiconductor Physics Laboratory Co. Ltd., Prielle Korn\'elia str. 2. H-1117 Budapest, Hungary}

\author{B. G. M\'{a}rkus}
\affiliation{Department of Physics, Budapest University of Technology and Economics
and MTA-BME Lend\"{u}let Spintronics Research Group (PROSPIN), P.O. Box 91, H-1521 Budapest, Hungary}

\author{A. J\'{a}nossy}
\affiliation{Budapest University of Technology and Economics, Institute of Physics and Condensed Matter Physics Research Group of the Hungarian Academy of Sciences, H-1521, Budapest P.O. Box 91, Hungary}

\author{F. Simon}
\thanks{Corresponding author: f.simon@eik.bme.hu}
\affiliation{Department of Physics, Budapest University of Technology and Economics
and MTA-BME Lend\"{u}let Spintronics Research Group (PROSPIN), P.O. Box 91, H-1521 Budapest, Hungary}

\author{G. Klupp}
\affiliation{Institute for Solid State Physics and Optics, Wigner
Research Centre for Physics, Hungarian Academy of Sciences, P.O. Box 49, H-1525 Budapest, Hungary}

\author{M. A. Gy\H{o}ri}
\affiliation{Institute for Solid State Physics and Optics, Wigner
Research Centre for Physics, Hungarian Academy of Sciences, P.O. Box 49, H-1525 Budapest, Hungary}

\author{K. Kamar\'{a}s}
\affiliation{Institute for Solid State Physics and Optics, Wigner
Research Centre for Physics, Hungarian Academy of Sciences, P.O. Box 49, H-1525 Budapest, Hungary}

\author{G. Magnani}
\affiliation{Dipartimento di Fisica e Scienze della Terra, Universit\`{a} degli Studi di Parma, Via G. Usberti 7/a, 43124 Parma, Italy}
\author{D. Pontiroli}
\affiliation{Dipartimento di Fisica e Scienze della Terra, Universit\`{a} degli Studi di Parma, Via G. Usberti 7/a, 43124 Parma, Italy}
\author{M. Ricc\`o}
\affiliation{Dipartimento di Fisica e Scienze della Terra, Universit\`{a} degli Studi di Parma, Via G. Usberti 7/a, 43124 Parma, Italy}

\begin{abstract}
The $10$~GHz microwave conductivity, $\sigma(T)$ and high field, $222$~GHz electron spin resonance (HF-ESR) of Li$_4$C$_{60}$ fulleride is measured in a wide temperature range. We suggest that the majority of ESR active sites and at least some of the charge carriers for $\sigma(T)$ are electrons bound to a small concentration of surplus or vacancy ions in the polymer phase. Both $\sigma(T)$ and the ESR line shape depend on ionic motion. A change of the activation energy of $\sigma(T)$ at $125$~K coincides with the onset of the ionic DC conductivity. The ESR line shape is determined mainly by Li ionic motion within octahedral voids below $150$~K. At higher temperatures, fluctuations due to ionic diffusion change the environment of defects from axial to effectively isotropic on the ESR time scale.
$\sigma(T)$ data up to $700$~K through the depolymerization transition confirm that the monomeric phase of Li$_4$C$_{60}$ is a metal.
\end{abstract}

\maketitle

\section{Introduction}

C$_{60}$ fullerene molecules form polymeric structures with unusual phenomena, such as e.g. metallic conductivity and antiferromagnetic order along chains \cite{JanossyPRL1994}. Neutral C$_{60}$ is polymerized by a light \cite{Rao93} or pressure \cite{Nunez95} induced [2+2] cycloaddition reaction. In these one- and two-dimensional polymeric structures four-membered carbon rings interconnect the fullerene molecules.
The polymerization of C$_{60}^{n-}$ anions is spontaneous in alkali intercalated fulleride salts. The structure of these polymers depends on the fulleride charge, $n$. $[2+2]$ cycloaddition is favored for low values of $n$, like in AC$_{60}$ (A $=$ K, Rb, Cs)\cite{pekker94}, while single interfullerene bonds are more stable \cite{oszi97} for $n\geq 3$ as in Na$_{4}$C$_{60}$ and Na$_{2}$AC$_{60}$ (A $=$ K, Rb)\cite{Bendele98b}. The size of the counter-ion also plays a role; e.g. in Na$_{2}$CsC$_{60}$ a moderate pressure is needed to stabilize the polymeric structure \cite{margadonna99}.

Charge storage applications of Li intercalated carbonaceous compounds were proposed as early as 1976 (Ref. \onlinecite{Besenhard1976}). In this respect, the Li$_{4}$C$_{60}$ fulleride polymer is of special interest. In the 2D polymeric layers of this compound, fullerenes are connected by single bonds along one direction and by $[2+2]$ cycloaddition in the other\cite{margadonna04, Ricco05, rols2015}. Li$_{4}$C$_{60}$ is a superionic conductor\cite{Li4PRL} with a high ionic conductivity {\color{black}($0.01$~S/cm at $300$~K)}. The ionic conductivity of Li$_{4}$C$_{60}$ is intrinsic and it arises from the special structure; at low temperatures half of the Li ions have an unoccupied neighbor site which is easily occupied at higher temperatures \cite{Li4PRL}.

Recently Mg$_2$C$_{60}$, a fulleride electronically and structurally similar to Li$_{4}$C$_{60}$, was also found to be an ionic conductor\cite{PontiroliCarbon2013}.
In the case of alkali fullerides, ionic conductivity is limited to fullerenes with small alkali metal ions, as the small trigonal aperture connecting interstitial sites within the polymeric framework hinders ionic diffusion. In fulleride crystals with larger alkali metal ions, diffusion requires a reorganization of the fullerene molecular positions. Larger alkali metal ions rearrange or diffuse only at phase transitions or at high temperatures, e.g. above $400$ K in Na$_{2}$C$_{60}$ [Ref. \onlinecite{klupp06_2}]. {\color{black}The electronic properties under pressure and calorimetric measurements of Li, Na and K doped fullerides were carried out in detail by the group of Sundqvist \emph{et al.}\cite{Sundqvist2008,Yao2010, Sundqvist2011,Sundqvist2015,inaba2015}.} The ionic conductivity depends also on other factors than the barrier between sites; counter-intuitively, the DC conductivity of Li$_{4}$C$_{60}$ polymer increases under pressure.

Nuclear magnetic resonance (NMR) and electron spin resonance showed that Li$_{4}$C$_{60}$ polymer has a non-magnetic, insulating ground state \cite{Ricco05} while the high temperature monomer phase is metallic [Ref. \onlinecite{Ricco07}]. The frequency and temperature dependence of the electric conductivity provides a direct information on the ionic and electronic conductivities and on structural changes. In the Li$_{4}$C$_{60}$ polymer, the low frequency conductivity up to $1$~MHz is dominated by the ionic contribution\cite{Ricco05}, while the microwave conductivity is electronic. Microwave frequencies (typically $10$ GHz) are well above the characteristic frequency of Li$^{+}$ ion movement and far below the plasma edge in fulleride metals \cite{GunnRMP} (typically $0.5-1$~eV, $50-100$~THz); thus the microwave conductivity is dominated by electronic contributions. Unpaired electrons at Li vacancy and/or Li surplus sites are ESR active and are affected by Li ion diffusion. A complex behavior above $200$~K of the electron spin resonance at $9$~GHz was reported in a previous study \cite{Arcon08}. The high spectral resolution at $222.4$ GHz of this study allows to follow the onset of Li$^{+}$ ion diffusion at the ESR active defect sites.

Here, we present microwave conductivity and high frequency electron spin resonance (HF-ESR) measurements on Li$_{4}$C$_{60}$ in the $40-700$~K temperature range. We study the electronic properties of the polymeric and the monomeric phases with particular attention to the dynamics of electrons in the superionic phase and to the depolymerization process. We find that diffusion of Li$^{+}$ ions above $125$~K induces an activated electronic conductivity in polymeric Li$_{4}$C$_{60}$. Ion diffusion also explains changes in the HF-ESR spectrum in the same temperature range. We trace the depolymerization process up to $700$ K and confirm that the high temperature monomer phase is a good conductor, in accordance with previously reported results\cite{Ricco07}.

\section{Experimental}

Li$_{4}$C$_{60}$ samples were prepared as described previously \cite{margadonna04}. A careful X-ray and NMR characterization confirmed the polymeric structure and the stoichiometry Li$_x$C$_{60}$ with $x = 4$, similarly to the samples investigated in Refs. \onlinecite{Li4PRL} and \onlinecite{Ricco07}. For infrared (IR) measurements, the sample was pressed in an Ar filled glove box into KBr pellets. Li and Na doped species were measured in the glove box with a Bruker Alpha spectrometer, K$_4$C$_{60}$ was measured in an air tight sample holder in a Bruker IFS 66v spectrometer at room temperature with a resolution of $2$~cm$^{-1}$.

Powder samples of Li$_4$C$_{60}$ were sealed in quartz tubes under $200$~mbar He for the microwave conductivity and ESR measurements. Microwave conductivity was measured using the cavity perturbation technique which is well suited for air sensitive powder samples and thus allows the study of alkali fulleride compounds \cite{bommeli95}. This method is based on the measurement of changes in the quality factor, $Q$, of a microwave cavity arising from microwave eddy currents in the sample\cite{nebendahl01}. The $10$~GHz TE${011}$ copper cavity has an unloaded quality factor of $Q_{0}\approx 10,000$. A nitrogen gas flow quartz cryostat allowed to vary the sample temperature between $130$~K and $700$~K while keeping the cavity temperature at $290$~K. Measurements in the $40-300$ K temperature range were performed in a similar cavity placed in a liquid He cryostat.
The powder samples were placed at the center of the cavity where in highly conducting samples the electric field is not excluded by depolarization effects.
The grain size of the sample was smaller than the microwave penetration depth and the conductivity is proportional to $1/Q-1/Q_{0}$ [Ref. \onlinecite{Klein1993}] where $Q$ and $Q_{0}$ are the quality factors of the cavity with and without the sample, respectively \cite{buravov71}. Only  relative variations of the conductivity are measured as the proportionality factor between microwave losses and conductivity depends on the unknown grain size distribution.

High frequency electron spin resonance spectra were recorded with a home-built spectrometer \cite{JanossyHFESR_APL2011} operating at $222.4$~GHz (corresponding to $7.93$~T for a $g$-factor of $g=2$). The maximum output power of the microwave source is $46$~mW. Microwave radiation is transmitted by a quasi-optical bridge and is detected with an InSb detector operating at $4.2$ K. The spectrometer has a sensitivity of about $3\times 10^{10}$ spins /($10^{-4}$~T$\sqrt{\text{Hz}}$). The resolution of ESR lines determined by the homogeneity of the magnet is $0.05$~mT.

\section{Results and discussion}
\subsection{Infrared spectroscopy}
\label{sub:IR}

\begin{figure}[h!]
\includegraphics[width=1\hsize]{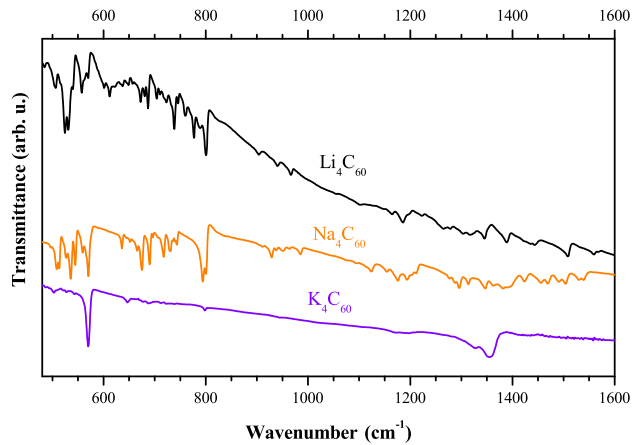}
\caption{Room-temperature infrared spectrum of Li$_4$C$_{60}$ compared to that of Na$_4$C$_{60}$ and
K$_4$C$_{60}$, shifted and scaled for clarity. Li$_4$C$_{60}$ and Na$_4$C$_{60}$ have similar IR spectra with a mode around $800~\text{cm}^{-1}$ signaling single bonds in the polymer, whereas the IR spectrum of monomeric K$_4$C$_{60}$ is markedly different.}
\label{fig:ir}
\end{figure}

Fig.~\ref{fig:ir} shows the infrared (IR) spectrum of Li$_4$C$_{60}$ and for comparison, the spectra of Na$_4$C$_{60}$ and
K$_4$C$_{60}$, where the charge state of the fullerene is also C$_{60}^{4-}$. The IR spectrum of Li$_4$C$_{60}$ resembles that of Na$_4$C$_{60}$, in agreement with the polymeric nature of both compounds. Na$_4$C$_{60}$ is a two-dimensional polymer with single bonds \cite{oszlanyi97}. Polymer formation significantly distorts the C$_{60}$ ball \cite{long00} {\color{black} and the original icosahedral symmetry of C$_{60}$} is lowered to $C_{\text{2h}}$ in Li$_4$C$_{60}$ [Ref. \onlinecite{margadonna04}] and $C_{\text{i}}$ in Na$_4$C$_{60}$ [Ref. \onlinecite{oszlanyi97}]. This results in the large number of IR active modes in contrast to the four allowed IR modes of C$_{60}$. On the other hand, the larger size of the alkali ion hinders the polymer formation in K$_4$C$_{60}$. {\color{black} Although there is a Jahn-Teller distortion \cite{klupp06}} even in this case, the distortion of the C$_{60}$ ball is smaller than in polymers and the IR active modes emerging from the lower symmetry are weak.

The strong band at $800$~cm$^{-1}$ in Li$_4$C$_{60}$ and Na$_4$C$_{60}$ arises from a single bond between fullerenes \cite{quintavalle08} and is not an intramolecular mode. Thus the infrared spectra confirm that single bonds are present in the Li$_4$C$_{60}$ polymer as determined by the structural characterization \cite{margadonna04}.

Infrared spectroscopy can in general provide information about the conductivity of materials. In the case of powders in KBr pellets, the effects of light scattering obscure the exact shape of the free-carrier (Drude) absorption, but the signs of metallic character are a strong background absorption and the change of the vibrational bands from Lorentzian to Fano shape\cite{zadik2015,kamaras2014}. We do not observe any of these effects here, which confirms the absence of such additional spectral weight in the IR spectrum of Li$_4$C$_{60}$ at room temperature. This invokes that the electronic conductivity is negligible in agreement with Ref. \onlinecite{Li4PRL}. Due to the large mass of the charge carrying ions in fulleride compounds, the Drude peak corresponding to the ionic conductivity is well below the usually accessible frequency of $100$ cm$^{-1}$.

\subsection{Electronic and ionic conductivities in the polymer phase}

\begin{figure}[h!]
\centerline{\includegraphics[width=1\hsize]{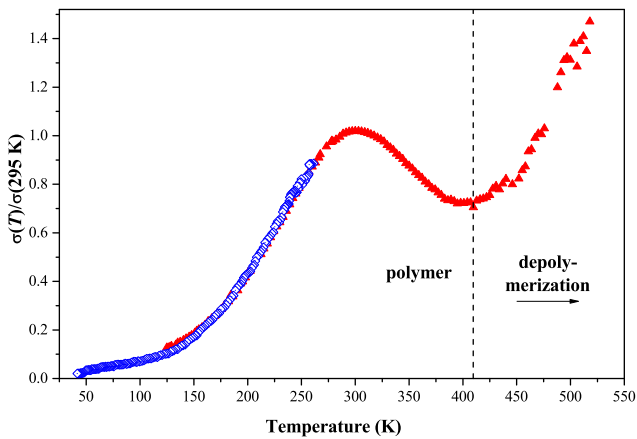}}
\caption{Temperature dependence of the $10$~GHz microwave conductivity of Li$_{4}$C$_{60}$ normalized at $295$~K. Red triangles: heating (heating rate $0.2$~K/s), blue dotted squares: cooling. Unpaired electron concentration increases above $300$~K, depolymerization is significant above $400$~K. }
\label{fig:mwpolimer}
\end{figure}

Fig.~\ref{fig:mwpolimer} shows the temperature dependence of the $10$~GHz microwave conductivity, $\sigma(T)$, of Li$_{4}$C$_{60}$ normalized to the room temperature value. Data above $260$~K are for increasing temperature only. Below $260$~K, data taken in heating and cooling are indistinguishable. The conductivity was too small to measure the microwave loss below $40$~K. At low temperature, $\sigma (T)$ increases with increasing $T$; it has a maximum around $300$~K and a minimum at $410$~K. Thermal cycles around the polymerization temperature show that the data in Fig.~\ref{fig:mwpolimer} below $400$~K corresponds to the polymeric phase. The subsequent rapid increase arises from the onset of depolymerization and is discussed in Sec.~\ref{sub:monomer}.

\begin{figure}[h!]
\centerline{\includegraphics[width=1\hsize]{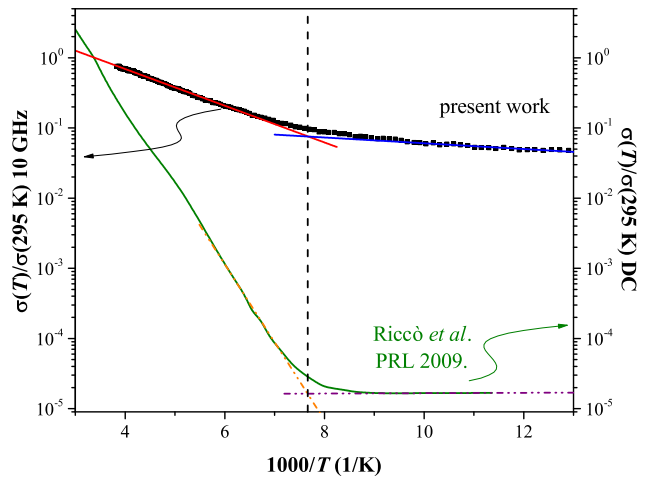}}
\caption{Arrhenius plot of the $10$~GHz (electronic) conductivity in Li$_4$C$_{60}$ (filled squares, this work) and the DC (ionic) conductivity (solid curve, Ref.~\onlinecite{Li4PRL}). The conductivities are normalized to their respective $295$~K values. The electronic conductivity is larger than the ionic, however, its absolute value is not known. Straight lines are fits to the data. Note the marked upturn in both electronic and ionic conductivities at about $125$ K, which is attributed to the onset of Li$^{+}$ diffusion.}
\label{fig:gaps}
\end{figure}

Fig. \ref{fig:gaps} compares the low temperature microwave conductivity with the DC conductivity data of Ref.~\onlinecite{Li4PRL} using an Arrhenius presentation. An activated behavior $\sigma(T) = \sigma_0 \mathrm{e}^{-\Delta / T}$ with different activation energies, $\Delta$, above and below $125$~K fits well the microwave conductivity data. The corresponding parameters are summarized in Table \ref{Table}. The activation energy of the DC ionic conductivity is much larger than for the microwave conductivity. Although the absolute value of the microwave conductivity is not known, it is certainly much larger than the DC conductivity at $125$~K, the onset of ionic diffusion and most likely remains larger up to $300$~K.

\begin{table}
\caption{The parameters used to fit the activated microwave conductivity of Li$_4$C$_{60}$ for the $T<125~\text{K}$ and $T>125~\text{K}$ regimes.}
\begin{tabular*}{1\columnwidth}{@{\extracolsep{\fill}}ccc}
 \hline
 \hline
 Regime & $\sigma_0$ (arb. u.) & $\Delta~(\textrm{K})$ \\
 \hline
 $T>125~\text{K}$ & $8.7(1) \times 10^{-4}$ & $603(4)$ \\
 $T<125~\text{K}$ & $1.7(1) \times 10^{-5}$ & $90(6)$ \\
 \hline
 \hline
\label{Table}
\end{tabular*}
\end{table}

The different temperature dependence of the DC and microwave conductivities suggest these have different origins.
In our view\cite{Li4PRL}, above the $125$~K onset of Li$^{+}$ ionic diffusion to about ambient temperatures the DC conductivity arises mainly from the ionic conductivity, $\sigma_{\text{ion}}$ involving all Li ions. (Sundqvist \emph{et al.} \cite{Sundqvist2015} raised, however some doubts about this). On the other hand, we suggest that the microwave conductivity is the electronic conductivity $\sigma_{\text{el}}$ at $10$~GHz associated with a small concentration of charged defects.

There is ample evidence for Li ionic diffusion at ambient temperatures and below\cite{Arcon08,Li4PRL}. It was observed by the motional narrowing of the $^{7}$Li NMR line and, as explained in Sec.~\ref{sub:ESR}, the narrowing of the ESR line above $125$~K is also well understood by Li ionic diffusion. The strong temperature and frequency dependence of the DC conductivity supports the dominant role of the ionic contribution\cite{Cattaneo2016}.

The microwave conductivity has an electronic origin; the ionic conduction is negligible at $10$~GHz. Except for very low frequencies, the ionic conductivity decreases strongly with frequency. An extrapolation of the low frequency data at $246$~K and below shows that $\sigma_{\text{ion}}$ at $10^{10}$ Hz is well below the sensitivity of the microwave cavity conductivity measurement technique. The total conductivity (i.e. ionic plus electronic) as a function of frequency has a temperature dependent minimum somewhere between $10^6$ and $10^9$~Hz. From DC to frequencies of the order of $10^6$~Hz the conductivity is dominated by the ionic contribution while at higher frequencies the conductivity is due to electrons bound to charged defects.

According to the IR experiment (Sec.~\ref{sub:IR}), the material is essentially an electronic insulator at ambient temperatures; the electronic band gap is large and most electrons do not contribute to the conductivity. We suggest that the electronic conductivity is associated with a small concentration of electrons trapped at defects of the lattice. Li vacancy or Li surplus sites are obvious candidates; for these sites the microwave conductivity and ESR originate from the same electrons. As explained in Sec. \ref{sub:ESR}, the unpaired defect electrons are confined to well defined states at octahedral voids below $125$~K. The associated microwave electronic conductivity is due to electron hopping between states in the vicinity of the charged defects. {\color{black}Calorimetric measurements also support this proposal. A contribution to the specific heat was attributed to Li$^{+}$ motion within octahedral voids from temperatures as low as $2$~K [Ref. \onlinecite{inaba2015}].} The disorder induced by Li ionic diffusion above $125$ K creates new electronic states and allows electronic diffusion of the small concentration of defect electrons to larger distances. This explains the stronger increase of conductivity above the onset of Li ionic diffusion at $125$~K.

{\color{black}Finally, we note that one may associate the slowly hopping localized electronic states giving rise to the microwave conductivity of the polymeric phase to small-polarons\cite{Kittel, solyom2, Holstein59, Jame97}. Li$_{4}$C$_{60}$ is an ionic salt with an insulator ground state. According to HF-ESR results (discussed in Section \ref{sub:ESR}), at low temperatures the excited states are confined to the vicinity of octahedral voids. These localized excitations diffuse slowly as temperature is raised. The small-polaron motion is a thermally activated process, which agrees well with our findings shown in Fig. \ref{fig:gaps}. The concentration of mobile electrons is small and small-polaron conduction in Li$_{4}$C$_{60}$ contributes significantly to the conductivity only at high frequencies, where the ionic contribution is negligible. At microwave frequencies, the electronic (polaronic) conduction determines the conductivity.}  The presence of polarons can also be inferred from HREELS measurements indicating coupling of electrons to low-energy (alkali metal or intermolecular) phonons [Ref. \onlinecite{macovez2008}].

\subsection{Narrowing of the ESR spectrum by Li diffusion}
\label{sub:ESR}

\begin{figure}[h!]
\centerline{\includegraphics[width=1\hsize]{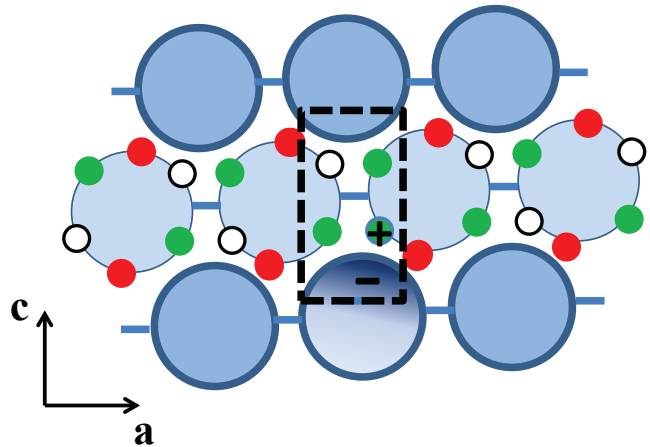}}
\caption{Schematics of the suggested ESR active defect. Large blue circles are polymeric fulleride ions. Dark and light molecules are in the front and back, respectively. Green (red) circles are Li$^{+}$ ions in octahedral (tetragonal) sites. In stoichiometric Li$_{4}$C$_{60}$, octahedral voids are occupied by $2$ Li$^{+}$ ions and have $2$ unoccupied sites. The unpaired electron is bound to the void occupied by $3$ Li$^{+}$ ions marked by a dashed line contour.}
\label{fig:defect}
\end{figure}

The ESR spectrum arises from defects of the polymeric matrix or from unpaired electrons in the polymeric structure bound to charges arising from a slight off-stoichiometry of the Li concentration. The ESR active defect concentration is one $S=1/2$ spin per hundred C$_{60}$ molecules \cite{Arcon08}. As explained below, at low temperatures the large majority of the ESR active sites have the same environment and we suggest that the ESR arises mostly from off-stoichiometry. Fig. \ref{fig:defect} is a schematic view of a charged ESR active site bound to a surplus Li$^{+}$ ion in an octahedral void. The resonance frequency of the ESR, characterized by the $g$-factor, is sensitive to small variations in the electronic configuration surrounding defects. The anisotropic $g$-shift depends on the spin-orbit interaction and the crystal field in an environment with lower than cubic symmetry and is very small in materials composed of light atoms\cite{FazekasBook}. The previous ESR study \cite{Arcon08} at $35$ GHz observed the deviation from cubic symmetry as a splitting of the ESR signal. The present higher resolution $222$ GHz ESR work follows the evolution of the environment of ESR active defects in detail. We find that the majority of defects evolve from an axial symmetry configuration at low temperatures to an effectively isotropic configuration at ambient temperatures.

\begin{figure}[h!]
\centerline{\includegraphics[width=1\hsize]{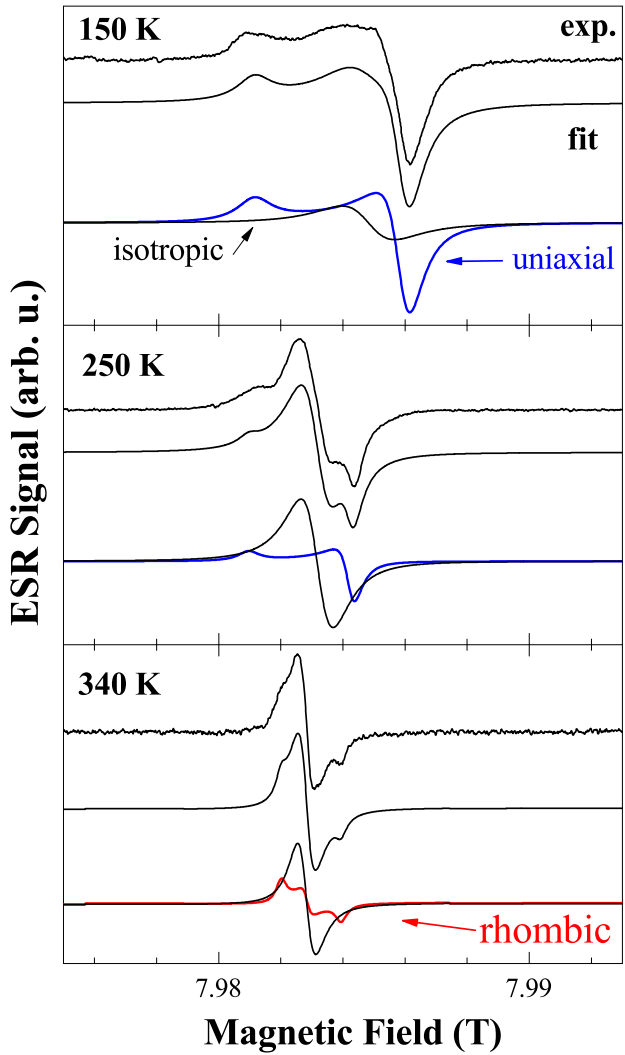}}
\caption{Temperature evolution of the $222$ GHz ESR spectrum of Li$_{4}$C$_{60}$. The measured experimental (exp.) and the fitted (fit) spectra are shown for each temperature. Below $290$ K, the spectra consist of an isotropic (black) and a uniaxially anisotropic $g$-factor distribution (blue) component. Above $290$ K, the anisotropic component is better described by a rhombic $g$-factor anisotropy (red).}
\label{fig:ESR1}
\end{figure}

Fig. \ref{fig:ESR1} displays the temperature evolution of the ESR spectrum of as-prepared Li$_{4}$C$_{60}$.
The spectrum is the superposition of `anisotropic' and `isotropic' components. The anisotropic component is a powder spectrum broadened by the $g$-factor anisotropy in the magnetic field. The line shape is characteristic of a uniaxial $g$-factor anisotropy (blue curve in Fig. \ref{fig:ESR1}) \cite{SlichterBook,AthertonBook}. The width of the anisotropic component decreases with temperature. Above $280$~K only a small intensity anisotropic line with a different line shape remains that possibly reflects a lower, rhombic, symmetry. The decrease of the line width of the anisotropic component is characteristic of a fluctuating environment in which the average $g$-factor anisotropy is reduced but remains finite. The isotropic component is a single line with a Lorentzian line-shape (solid line in Fig. \ref{fig:ESR1}), characteristic of sites with a cubic static environment or with a rapidly varying environment that averages the $g$-factor anisotropy to zero.

\begin{figure}[h!]
\centerline{\includegraphics[width=1\hsize]{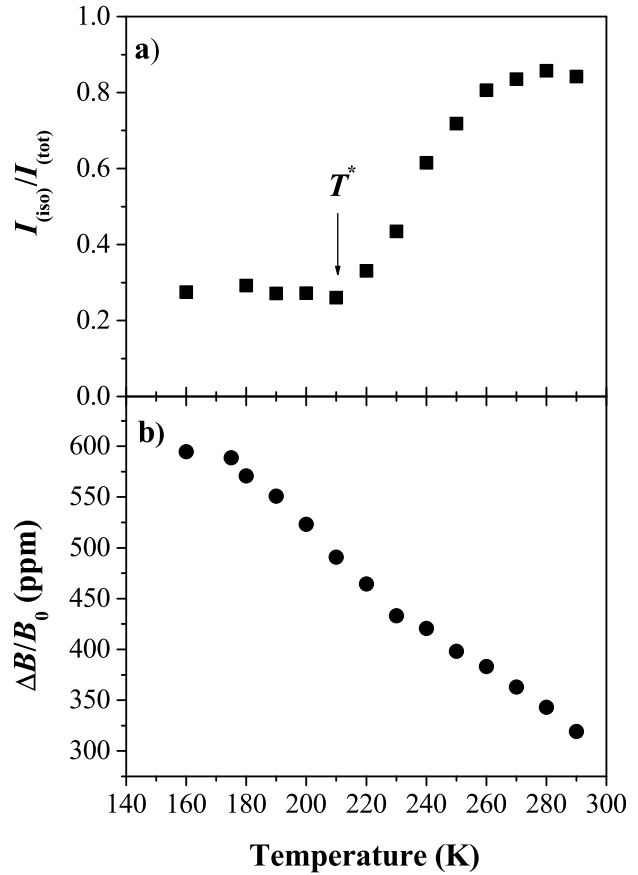}}
\caption{a) Temperature dependence of ESR intensity of the isotropic component divided by the total one , $I_{\text{(iso)}}$/$I_{\text{(tot)}}$. $T^{*}$ indicates the onset of Li$^{+}$ ion motion.\cite{Arcon08} b) Temperature dependence of the $g$-factor anisotropy, $\Delta B/B_{\text{0}}$, of the ESR line measured at the same frequency ($222$~GHz).}
\label{fig:ESR2}
\end{figure}

Fig. \ref{fig:ESR2} displays the detailed variation of the relative intensity and the line width for the anisotropic component.
The intensity of the isotropic component relative to the total intensity of the spectrum, $I_{\text{(iso)}}$/$I_{\text{(tot)}}$ is constant at low temperatures and increases in a step-like fashion between $210$ and $240$~K
(Fig. \ref{fig:ESR2}a). The spectrum is dominated by the anisotropic component between $160$ and $210$~K and by the isotropic component above $260$~K.
The $g$-factor anisotropy of the anisotropic component decreases gradually with increasing temperature from $6.4$~mT at $160$~K to $2.6$~mT at $290$~K (Fig. \ref{fig:ESR2}b).
The defect concentration changes less than a factor of $2$ below $280$~K since the spin susceptibility, measured by the ESR spectrum intensity, $I_{\text{(tot)}}$, depends little on temperature\cite{Arcon08} between $150$~K and $280$~K. The long electronic spin-lattice relaxation time prevents reliable intensity measurements at lower temperatures. The spin susceptibility increases rapidly above $300$~K, indicating the formation of new ESR active defects.

We interpret the unusual temperature dependence of the ESR spectrum as a consequence of the fluctuation of the $g$-factor tensor of unpaired localized electrons arising from Li$^{+}$ motion.
At low temperatures, Li ions in tetrahedral sites are static. Li ions in octahedral voids are hopping between a few possible sites and remain within voids for long times. We assume that the anisotropic component arises from unpaired electrons bound to octahedral voids with a third Li ion in addition to the two Li ions of the majority sites. (Electron holes bound to Li vacancies is another possibility). The charge of the Li ion nearest to the unpaired electron is the most important factor for the deviation from cubic symmetry of the crystal field at the unbound electron. The orientations and principal values of the $g$-factor tensor of a given ESR active site fluctuates between a few well defined values due to hopping of Li ions within the voids but within the ESR time scale the anisotropy is not averaged to zero.
The relevant ESR time scale $\tau_{\text{ESR}} = 5 \times 10^{-9}~\text{s}$ is given by the low temperature width of the anisotropic line.
Non-identified sites, representing $20$ percent of the localized unpaired electron defects, give a narrow, isotropic line at low temperatures.

At higher temperatures, the density of unpaired electronic states increases gradually with the increasing disorder and this decreases the effective $g$-factor anisotropy. At temperatures above $210$~K, the fluctuations of the $g$-factor due to diffusion of Li ions between different voids in the polymeric structure becomes important and the line shape changes in a qualitative way. Electrons bound to (or in the vicinity of) the diffusing surplus ions experience a rapidly changing environment between many more states than experienced at lower temperatures. Diffusion involves all Li ions between voids according to Ref. \onlinecite{Li4PRL}. Each C$_{60}$ molecule of the polymer is surrounded by $8$ equidistant tetrahedral and $6$ octahedral voids. Diffusion in and out the large number of sites averages the $g$-factor anisotropy and contributes the to a narrow, on-the-average isotropic line intensity. The concentration of unpaired electrons at isotropic sites increases rapidly between $210$ and $260$~K but the total defect concentration does not change. Above $260$~K Li ion hopping between octahedral voids determines the ESR lineshape. At $340$~K only a very small intensity anisotropic line is observed around the isotropic line.

The motion of Li ions was observed in a broad temperature range by the narrowing of the $^7$Li NMR spectrum \cite{Arcon08}. The temperature dependence of the NMR and the ESR spectra are qualitatively similar. The NMR spectrum at low temperatures has an "anisotropic" broad component due to the distribution of electric field gradients (EFG) at octahedral Li sites. Motion of the Li ions reduces the time-averaged EFG measured by the NMR line width of the broad component. Like for the anisotropic ESR, there are two temperature ranges. The NMR line narrows moderately with temperature up to $190$ K. At higher temperatures the EFG decreases rapidly in a step-like fashion and the narrow line at high temperature is characteristic of an isotropic environment on the NMR time scale.

Microwave conductivity, $^7$Li NMR, and ESR are sensitive to both intra- and inter-void hopping but at different time scales. The DC ionic conductivity senses only ion diffusion between voids.
The onset of the increasing isotropic ESR signal intensity (shown with an arrow in Fig. \ref{fig:ESR2}) is  thus associated with the onset of Li$^{+}$ ion hopping between voids around $210$ K. This temperature is slightly higher than the onset temperature of the rapid narrowing of the anisotropic $^7$Li NMR line. The DC ionic conductivity is more sensitive to the onset of hopping between voids and is observed from much lower temperatures. As discussed earlier, the microwave electronic conductivity is also influenced by the ionic motion.

Narrowing of the ESR spectra allows to estimate the correlation time for Li diffusion between voids as it is of the order of the correlation time, $\tau_{\text{m}}$ of the $g$-factor fluctuation \cite{SlichterBook}.
\begin{equation}
\Delta \omega_{\text{iso}} (T) = \Delta \omega^2 (0)~\cdot \tau_{\text{m}},
\label{motional_narrowing}
\end{equation}
where $\Delta \omega (0)$ is the width of the static low temperature anisotropic spectrum (in angular frequency units) and $\omega_{\text{iso}} (T)$ is the width of the high temperature isotropic line.
Inserting $\Delta \omega(0)/\gamma=6.4\,\text{mT}$ ($\gamma/2\pi=28.0$~GHz/T is the electron gyromagnetic ratio) and $\Delta \omega_{\text{iso}} (300~\text{K})$ = $0.7$~mT, we find $\tau_{\text{m}}= 10^{-10}~\text{s}$.

\subsection{Depolymerization and metallic conductivity in the monomeric phase}
\label{sub:monomer}

\begin{figure}[h!]
\centerline{\includegraphics[width=1\hsize]{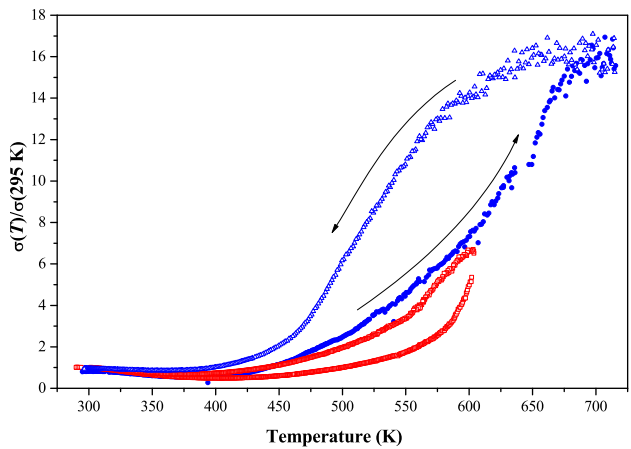}}
\caption{ Depolymerization - polymerization cycles of Li$_{4}$C$_{60}$ measured by the $10$~GHz microwave conductivity (normalized at $295$~K). Full cycle: heating (blue dots) and cooling (blue triangles) rates $\approx 0.1$~K/s. Partial cycle (red squares): cooling and heating rates $\approx 0.07$~K/s.}
\label{fig:depolimer}
\end{figure}

X-ray diffraction at high temperatures shows a depolymerization of the low temperature monoclinic polymeric phase of Li$_{4}$C$_{60}$ into a cubic monomeric phase \cite{Ricco07}. The depolymerization is hysteretic; in the XRD study a mixture of monoclinic and cubic phases appears between 470 and 600 K. ESR, NMR, and Raman spectroscopy data suggests that the cubic monomer phase is metallic \cite{Ricco07,Arcon08}.  We confirm in this work the metallic character of the high temperature phase by a direct measurement of the electronic conductivity. The structural phase transition between the electronic insulating low temperature polymeric phase and the high temperature metallic monomeric Li$_{4}$C$_{60}$ phases is followed in the $10$ GHz microwave conductivity. Fig. \ref{fig:depolimer} displays two examples of the several temperature cycles performed. The heating and cooling cycle between $300$ and $710$ K is between the polymeric phase, well below the depolymerization temperature, to above $670$~K where the material is homogeneously monomeric. In the other cycle displayed, a large part of the material remains polymeric up to $600$~K.

The $10$~GHz conductivity has a local maximum at $300$~K and a minimum at about $400$ K in all heating curves with various heating rates and history. The anomaly differs little between the heating and cooling parts of the $300$ and $600$~K cycle where the transition to the monomer phase is only partial. Thus the maximum in the conductivity at $300$~K is a feature of the polymeric phase. We suggest that the anomalous decrease in the $10$ GHz electronic conductivity between $300$ K and $400$~K is due to the reversible breaking of some of the polymeric bonds.
	
Above $400$~K, the conductivity increases rapidly in the heating cycle as metallic domains of the monomeric phase are formed. The hysteresis of the $10$ GHz conductivity places the polymerization temperature at about 550~K in agreement with the X-ray diffraction data \cite{Ricco07}. For the cycle shown in Fig. \ref{fig:depolimer}, hysteresis is very small above $670$~K. The conductivity is approximately temperature independent above this temperature and an under-cooling effect is observed well below $670$ K. The monomer phase is metallic with more than an order of magnitude higher conductivity than in the polymeric phase at ambient temperature. As expected for a metal, the spin susceptibility of the monomer phase measured by ESR is high and approximately temperature independent \cite{Arcon08}. (This measurement was, however in cooling from $620$~K of an incompletely polymerized sample).

\section{Conclusions}

We studied the Li$_4$C$_{60}$ fulleride using microwave conductivity and high field electron spin resonance spectroscopy in a wide temperature range.
Low temperature microwave conductivity changes near the temperature where ionic conduction starts, which indicates a connection between the electronic and ionic conductions. HF-ESR  measurements demonstrate that the Li$^{+}$ ion dynamics deeply influences the electronic configuration of the paramagnetic centers in the polymer superionic phase. For $T<200$~K, the ions are static and localize the electrons which give rise to an ESR line with $g$-factor anisotropy. When Li$^{+}$ ions motion is activated, localized electrons start to diffuse with different velocities. The hopping of localized electrons gives the most important contribution to the conductivity of the polymer phase in the microwave frequency range where the ionic conductivity is negligible. Near 300~K, the conductivity has a maximum due to the onset of defect formation in the polymeric phase. The complex dynamics of the electrons in the polymer phase also results in an unusual temperature dependence of the $g$-factor anisotropy. Microwave conductivity measurements confirm the metallic nature of the high temperature monomeric phase Li$_{4}$C$_{60}$ compound.

\section{Acknowledgments}
Work supported by the ERC Starting Grant No. ERC-259374-Sylo and OTKA 105691.


\end{document}